\documentclass[twocolumn,english,aps,prl,preprint,reprint,showpacs,longbibliography,showkeys,superscriptaddress]{revtex4-1}
\usepackage[T1]{fontenc}
\usepackage{lmodern}
\usepackage{color}
\usepackage{babel}
\usepackage{amsmath}
\usepackage{amssymb}
\usepackage{graphicx}

\RequirePackage[colorlinks=true,linkcolor=blue]{hyperref}
\usepackage{dsfont}
\usepackage{cleveref}
\usepackage{natbib}
\usepackage{bm}
\usepackage[bottom]{footmisc}
 
\makeatletter
\usepackage{amsfonts}
\usepackage{babel}
\usepackage{braket}
\usepackage{verbatim}
\usepackage{amstext}
\makeatother

\begin{document}

\title{Electrical control and decoherence of the flopping-mode spin qubit}
\title{Electric-field control and noise protection of the flopping-mode spin qubit}

\author{M. Benito}
\address{Department of Physics, University of Konstanz, D-78457 Konstanz, Germany}
\author{X. Croot}
\address{Department of Physics, Princeton University, Princeton, New Jersey 08544, USA}
\author{C. Adelsberger}
\address{Department of Physics, University of Konstanz, D-78457 Konstanz, Germany}
\author{S. Putz}
\thanks{Present address:  Vienna Center for Quantum Science and Technology, Universit\"at Wien, 
1090 Vienna, Austria.}
\author{X. Mi}
\thanks{Present address: Google Inc., Santa Barbara, California 93117, USA.}
\author{J. R. Petta}\address{Department of Physics, Princeton University, Princeton, New Jersey 08544, USA}
\author{Guido Burkard}\address{Department of Physics, University of Konstanz, D-78457
  Konstanz, Germany}

\date{\today}

\begin{abstract}
We propose and analyze a novel ``flopping-mode'' mechanism for electric dipole spin resonance
based on the delocalization of a single electron across a double quantum dot confinement
potential.
Delocalization of the charge maximizes the electronic dipole moment compared to
the conventional single dot spin resonance configuration.
%
We present a theoretical investigation
of the flopping-mode spin qubit properties through the crossover from the double to the single
dot configuration by calculating effective spin Rabi frequencies and single-qubit gate fidelities.
The
flopping-mode regime optimizes the artificial spin-orbit effect generated by an external
micromagnet and draws on the existence of an externally controllable sweet spot, where the
coupling of the qubit to charge noise is highly suppressed.
We further analyze the sweet
spot behavior 
in the presence of a longitudinal magnetic field gradient,
which gives rise to a second
order sweet spot with reduced sensitivity to charge fluctuations.

\end{abstract}


\maketitle
\section{Introduction}
\label{sec:Introduction}

Control of individual electron spins is one of the cornerstones of spin-based quantum technology. 
Although 
standard
single-electron  spin resonance 
has been demonstrated ~\cite{Koppens2006}, there is a 
strong incentive 
to avoid the use of local oscillating magnetic fields since these are technically
demanding to generate at the nanoscale, hinder individual addressability, and  limit the Rabi frequency due to sample heating issues.
Electric dipole spin resonance (EDSR) techniques 
offer a more robust method to electrically control
the electron spin state.
Traditionally, successful implementations have used spin-orbit coupling ~\cite{Nowack2007},  hyperfine interaction \cite{Laird2007} and g-factor modulation ~\cite{Kato2003}.

\begin{figure}[]
\includegraphics[width=\columnwidth]{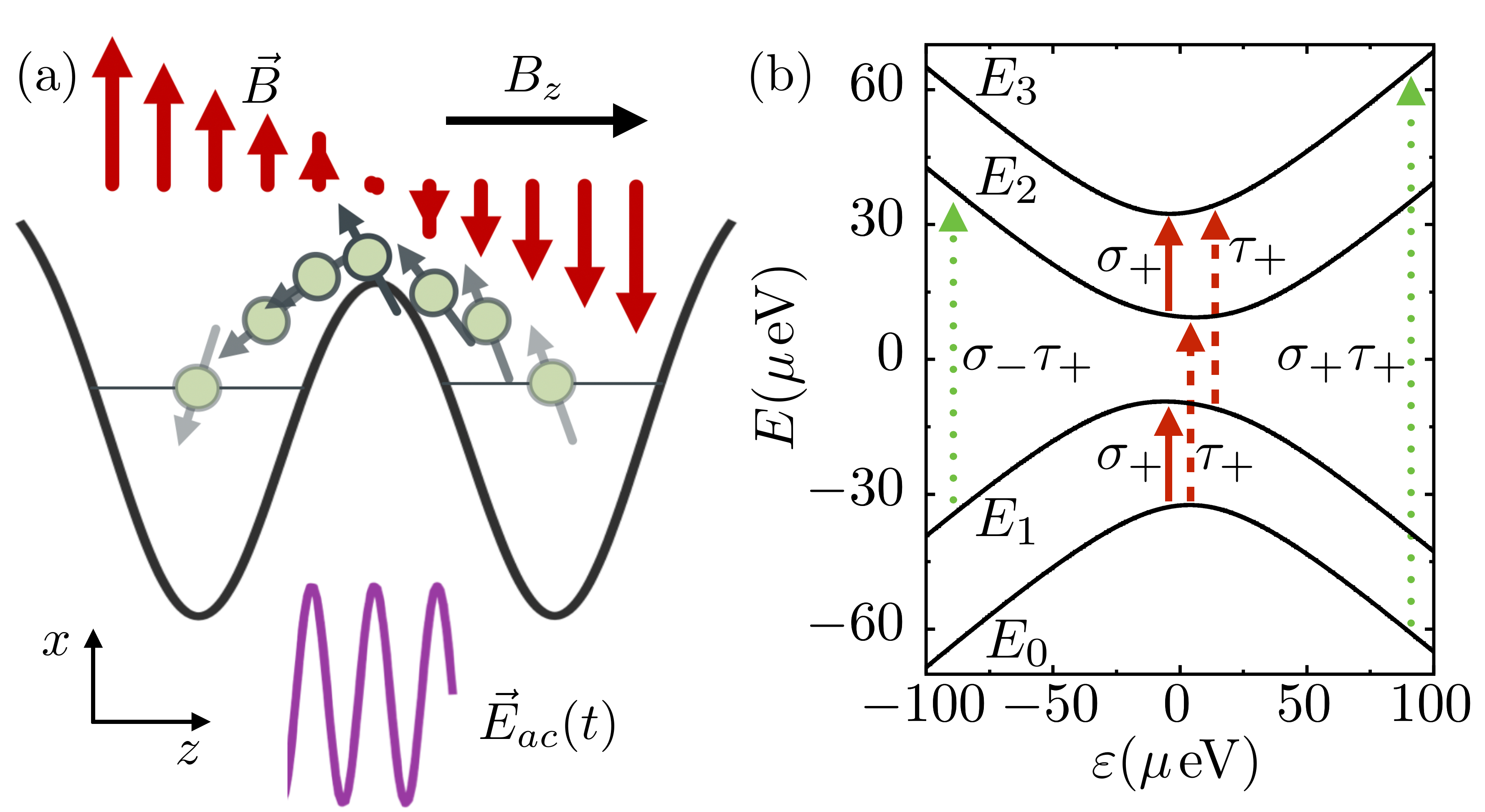}
\protect\caption{\label{fig:figure1} 
(a) Schematic illustration of the flopping-mode EDSR mechanism, where the spin of an electron (shown as green circles)  delocalized between two QDs  is driven via an electric field (purple line) in a magnetic field gradient (represented with red arrows).
(b) Energy levels $E_{0,\dots,3}$ of the Hamiltonian~\eqref{eq:total-H} as a function of the interdot detuning $\varepsilon$, calculated with $t_c = 20\, \mu\text{eV}$, $E_z = 24\, \mu\text{eV}$,  $g\mu_B b_x = 15\, \mu\text{eV}$, and $g\mu_B b_z = 4\, \mu\text{eV}$.
The asymmetry with respect to $\varepsilon$ is due to the longitudinal magnetic field gradient.
Around zero detuning, $|\varepsilon|\ll 2 t_c$,  the electron delocalizes across the
DQD, yielding a larger electric dipole moment $p$ compared to the single dot regime. 
The arrows represent the electrically addressable   spin (solid line), charge (dashed line) and spin-charge (dotted line) transitions. }
\end{figure}

The transition from GaAs to Si-based spin qubits has led to dramatic advances in the field of spin-based quantum computing. 
Site-selective single-qubit control~\cite{Veldhorst2014,Takeda2016,Yoneda2018}, two-qubit operations with high fidelity~\cite{Veldhorst2015,Zajac2016,Watson2018,Huang2018,Xue2018,Sigillito2019},  electron shuttling~\cite{Mills2018}, and strong coupling to microwave photons~\cite{Mi2018,Samkharadze2018}  have been demonstrated.
Recent demonstrations of strong spin-photon coupling have used double quantum dot (DQD) structures where the charge of one electron is 
delocalized between
both dots (``flopping-mode''; Fig.~\ref{fig:figure1}(a)), thus enhancing the coupling strength to the cavity electric field beyond the decoherence 
rate
 ~\cite{Mi2017b,Stockklauser2017,Bruhat2018} and
enabling the transfer of information between electron-spin  qubits  and 
microwave photons
~\cite{Mi2018,Samkharadze2018,Cubaynes2019}.
This suggests that the  manipulation of electron spins with classical electric fields  will also be efficient 
in the flopping-mode configuration.

The scalability of spin qubit processors hinges upon the use of resources that 
permit fast control  %
 without a 
significant degradation in coherence times.
The same properties that make silicon based QDs extremely attractive for quantum information processing make it challenging to use its intrinsic properties for electrical spin manipulation.
 Not only is the  hyperfine interaction to nuclear spins  largely reduced, but the intrinsic spin-orbit coupling for electrons in Si is very weak~\cite{Zwanenburg2013}.  
Recently, this weak effect combined with the rich valley physics in Si has been harnessed to achieve EDSR for single-electron spin qubits 
~\cite{Veldhorst2015b,Corna2018} and singlet-triplet qubits~\cite{Jock2018,Harvey-Collard2018}.
A more flexible solution applicable to any semiconductor is the mixing of orbital motion and spin via an externally imposed magnetic field gradient %
~\cite{Pioro-Ladriere2008,Kawakami2014,Yoneda2018}.
Beyond this effective spin-orbit effect, the control over the magnetic field profile 
  allows for selective addressing of spins placed in neighboring dots, since the resonance frequency varies spatially ~\cite{Pioro-Ladriere2008,Obata2010,Nadj-Perge2010,Yoneda2014,Noiri2016,Takeda2016,Ito2018}. 
Here we investigate  the effect of the micromagnet stray field on the coherence of the flopping-mode spin qubit.

In this work we envision the generation of single-electron spin rotations via a 
flopping-mode approach,
which benefits from the electron delocalization between two gate-defined tunnel coupled QDs~\cite{Hu2012}, 
and track its performance as    the electron is spatially localized in a single quantum dot (SQD).
The electron tunneling in such a double dot potential has a  large electric dipole moment, which is partially transferred to the spin via the magnetic field gradient induced by the stray field of a 
 micromagnet placed over the DQD, see Fig.~\ref{fig:figure1}(a). 
Moreover, due to the spatial  separation between the two QDs, obtaining a sizable 
magnetic field inhomogeneity,
with the resulting large effective spin orbit coupling,  becomes relatively easy.  A driving  field on one of the gate electrodes that shapes the QD  modulates the potential and allows full electrical spin control via EDSR.

The paper is organized as follows: In Sec.~\ref{sec:flopping-mode} we introduce the flopping-mode spin qubit and derive the Rabi frequency and the relevant relaxation and dephasing rates 
under the effect of a transverse magnetic field gradient
for the case of zero energy level detuning. 
In Sec.~\ref{sec:crossover}  we take into account the effect of a general detuning and analyze the electrical control of the flopping-mode spin qubit as a function of externally controllable parameters. In Sec.~\ref{sec:sweet-spots}  we investigate the  behavior of the flopping-mode spin qubit in the presence of a longitudinal magnetic field gradient and how this affects the working points with maximal single-qubit average gate fidelity. In Sec.~\ref{sec:conclusions} we summarize our results and conclude.

\section{Flopping-mode spin qubit}
\label{sec:flopping-mode}

An electron
trapped in a symmetric DQD, with zero energy level detuning $\epsilon = 0$ between the left (L)
and right (R) QDs will form bonding and antibonding charge states, which are separated by an
energy $2t_c$, where $t_c$ is the interdot tunnel coupling.
The transition dipole moment  between the  bonding and antibonding states, $|\mp\rangle=\left(|R\rangle\mp|L\rangle\right)/\sqrt{2}$, is proportional to the electronic charge $e$ and the distance between the two QDs $d$~\cite{Kim2014,Stockklauser2017,Bruhat2018,Mi2018}, therefore an electric field with amplitude  $E_{ac}$  at the position of the DQD can drive transitions with Rabi frequency $\Omega_c=edE_{ac}/\hbar$. 
Spins can be addressed via electric fields by splitting
the spin states via a homogeneous magnetic field, $B_z$, and inducing an inhomogeneous magnetic field   perpendicular to the  spin quantization axis, i.e., transverse ($\pm b_x$ in the left/right QD). 
We model the spin and charge dynamics with the Hamiltonian
\begin{equation}
H_0^{\varepsilon=0}=t_c\tilde{\tau}_z+\frac{E_z}{2}\tilde{\sigma}_z-\frac{g\mu_B b_x}{2}\tilde{\sigma}_x\tilde{\tau}_x \ , \label{eq:H0-eps0}
\end{equation}
where $\tilde{\tau}_{\alpha}$ and $\tilde{\sigma}_{\alpha}$ ($\alpha=x,y,z$) are the Pauli matrices in the charge ($|\pm\rangle$) and spin subspace, respectively,
 $E_z$ is the  Zeeman energy $E_z=g\mu_B B_z$, $g$ is the electronic g-factor  and $\mu_B$ the Bohr magneton.
The magnetic field gradient acts as an artificial spin-orbit interaction and hybridizes bonding and antibonding  states with opposite spin direction via the two spin-orbit mixing angles $\phi_{\pm}=\arctan{\left[g\mu_B b_x/(2t_c\pm E_z)\right]}$ ($\phi_{\pm}\in [0,\pi]$). 
As a consequence of this mixing, 
the electric dipole moment operator acquires off-diagonal matrix elements in the eigenbasis of Eq.~\eqref{eq:H0-eps0} which involve  spin-flip transitions~\cite{Benito2017,Benito2019}.
In particular, given the four eigenenergies 
$E_{0,\dots,3}$, with $2E_{3(2)}=-2E_{0(1)}=\sqrt{(2t_c\pm E_z)^2+(g\mu_B b_x)^2}$,
if $\tau$  denotes the two-level-system with energy splitting 
$E_{\tau}=E_2-E_0$ 
and $\sigma$ the one with splitting 
$E_{\sigma}=E_1-E_0$
(see Fig.~\ref{fig:figure1}(b)), the electric dipole moment operator reads 
\begin{align}
p &= e d \left[-\cos{\bar{\phi}}\tau_x+\sin{\bar{\phi}}\sigma_x\tau_z\right] \ , 
\label{eq:H0-diagonal}
\end{align}
where   $\bar{\phi}=\left(\phi_++\phi_-\right)/2$, and $\tau(\sigma)_{\alpha}$ ($\alpha=x,y,z$) are the Pauli matrices in the corresponding $\tau(\sigma)$ subspace.
This implies that the electric field can drive transitions between the ground state and  the first and second excited states with Rabi frequency 
$\Omega_{\sigma}=\Omega_c \sin{\bar{\phi}}$ and $\Omega_{\tau}=\Omega_c \cos{\bar{\phi}}$, respectively; see the center part of Fig.~\ref{fig:figure1}(b), where we have defined $2\tau_{\pm}=\tau_{x}\pm i \tau_{y}$ and $2\sigma_{\pm}=\sigma_{x}\pm i \sigma_{y}$. 

For $2t_c<E_z$ ($ 2t_c>E_z$), we define the spin qubit as  $s=\tau$ ($s=\sigma$), i.e., as the ground state and the second (first) excited state, with Rabi frequency $\Omega_s=\Omega_\tau$ ($\Omega_s=\Omega_\sigma$). %
If the transverse magnetic field  is small,  $g\mu_B b_x\ll |2t_c-E_z|$, the expansion to first order yields
\begin{equation}
\Omega_{s}=  2t_c g\mu_B b_x \Omega_c /|4t_c^2-E_z^2|+\mathcal{O}(b_x^3)  \label{eq:gc-expansion-eps0}
\end{equation}
for both $2t_c <E_z$  and $ 2t_c>E_z$.
For a very small (or very large) tunnel splitting, $2t_c$, the qubit is an almost pure spin qubit and it is hardly addressable electrically, while in the region $2t_c\approx E_z$ 
the spin-electric field coupling is maximal~\cite{Benito2017}
but the spin qubit coherence suffers to some extent from charge noise (see below).

The spin or charge character of the qubit will be reflected in the decoherence time. 
The spin-charge mixing mechanism also couples the spin to the phonons in the host material, therefore
the  
relaxation rates via phonon emission are 
$\gamma_{1,\sigma}=\gamma_{1,c} \sin^2(\bar{\phi})$ 
and $\gamma_{1,\tau}=\gamma_{1,c} \cos^2(\bar{\phi})$~\cite{Srinivasa2013}, respectively,
where we have introduced $\gamma_{1,c}$ as the  relaxation rate from the antibonding to the bonding state evaluated at the qubit energy.  
Since the  spin qubit energy is essentially given by the Zeeman splitting $E_z$ (weakly corrected by the spin-charge mixing), we can safely assume a constant value for $\gamma_{1,c}$, neglecting both oscillations of the form $\cos{(q d)}$ (q is the phonon quasimomentum) and polynomial dependences on the transition frequency~\cite{Brandes1999,Tahan2014,Raith2011,Hu2011,Kornich2018}.
The expansion to the lowest order in $b_x$ yields 
\begin{equation}
\gamma_{1,s}= \gamma_{1,c} \left[2t_c g\mu_B b_x/(4t_c^2-E_z^2)\right]^2+\mathcal{O}(b_x^4) \ ,
\end{equation}
where we can evaluate $\gamma_{1,c}$ at the Zeeman splitting energy $E_z$.
 In the symmetric configuration $\varepsilon=0$, pure dephasing  is strongly suppressed since the qubit is in a sweet spot protected to some extent from charge fluctuations~\cite{Vion2002,Petersson2010}.
Although the qubit energy splitting is first-order insensitive to electrical fluctuations in detuning $\varepsilon$,  
we  account here for  pure dephasing due to  second-order coupling to  charge fluctuations, 
which induces a Gaussian decay of coherences ($\propto e^{-(\gamma^{(2)}_{\phi,\sigma(\tau)} t)^2}$) with rates
 $\gamma^{(2)}_{\phi,\sigma}=(\gamma_{\phi}^2/E_{\sigma})\sin^2\bar{\phi}$ and $\gamma^{(2)}_{\phi,\tau}=(\gamma_{\phi}^2/E_{\tau}) \cos^2\bar{\phi}$, where  $\gamma_{\phi}$ is the magnitude of the low-frequency detuning charge fluctuations  (see Appendix ~\ref{sec:full-dipole}).
The expansion to the lowest order in $b_x$ yields 
\begin{equation}
\gamma_{\phi,s}^{(2)}=\frac{ \gamma_{\phi}^2}{E_z} \left[2t_c g\mu_B b_x/(4t_c^2-E_z^2)\right]^2+\mathcal{O}(b_x^4) \ .
\end{equation}

Note that far from the resonant point $2t_c\approx E_z$, other decoherence sources related to the spin, such as the hyperfine interaction with
nuclear spins, would start dominating the dephasing. The  dephasing  corresponding to quasistatic magnetic noise~\cite{Taylor2007,Chekhovich2013} with magnitude $\gamma_{M}$  is also quadratic, and the corresponding rates  are 
$\gamma_{M,\sigma}=\gamma_{M}(\cos{\phi_+}+\cos{\phi_-})/2$ and $\gamma_{M,\tau}=\gamma_{M} (\cos{\phi_+}-\cos{\phi_-})/2$ (see Appendix \ref{sec:magnetic-noise}).
 Therefore, to lowest order in $b_x$, the  spin qubit magnetic noise dephasing rate is 
\begin{equation}
\gamma_{M,s}= \gamma_{M} \left[1-\frac{(g\mu_B b_x)^2(4t_c^2+E_z^2)}{2(4t_c^2-E_z^2)^2}\right]+\mathcal{O}(b_x^4) \ .
\end{equation}

In this architecture the electric field can induce spin rotations with Rabi frequency $\Omega_s$.
%
We focus on the shortest single-qubit spin rotation ($X_{\pi}$ gate), performed in the gate time $t_g=\pi/\Omega_s$. Using a master equation with qubit relaxation and a noise term, we calculate  the average gate fidelity  (see Appendix \ref{sec:average-fidelity})
and average this result over a Gaussian distribution for the noise with standard deviation given by the total magnitude of the low-frequency  noise, $\text{Var}(\delta)=2\left({\gamma_{\phi,s}^{(2)}}^2+\gamma_{M,s}^2\right)$.
The optimal tunnel coupling value to achieve the best
single-qubit average gate fidelity
 depends on the relation between 
 the charge-induced dephasing  
 and the magnetic noise  (see Sec.~\ref{sec:crossover}).
Note that if the DQD is coupled to a microwave resonator the spin qubit couples also to the confined electric field and the Purcell effect opens another relaxation channel via photon emission.
Single-spin control  was demonstrated in Ref.~\cite{Mi2018}  in a detuned DQD configuration, where the spin-charge mixing, and therefore the coupling of the spin to the electric field is much weaker. In the following we analyze the crossover from a symmetric (DQD) to a far detuned  (SQD) configuration.


\section{Crossover from DQD to SQD}
\label{sec:crossover}

\begin{figure}
\includegraphics{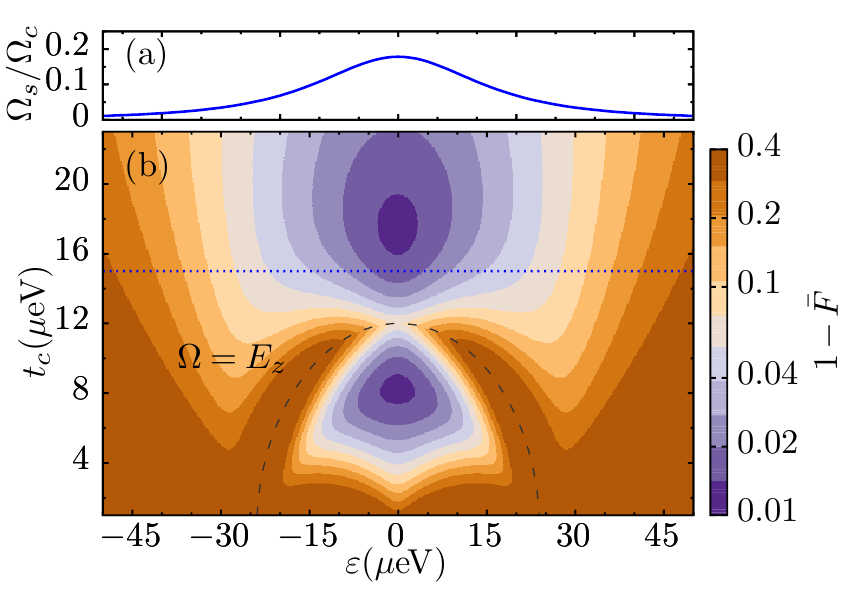}

\protect\caption{\label{fig:figure2} 
(a) Ratio between the spin Rabi frequency  $\Omega_s$  and the charge Rabi frequency $\Omega_c$ 
as a function of detuning $\varepsilon$ for $t_c=15\, \mu\text{eV}$. The spin Rabi frequency is maximized for $\varepsilon=0$.
(b) Single-qubit average gate infidelity
as a function of $\varepsilon$  and  $t_c$.
As expected,
$\bar{F}$ 
is symmetric about $\varepsilon = 0$, with the highest values  achieved at
$\varepsilon = 0$
and slightly away from the line with maximal spin-charge mixing, $\Omega=E_z$ (black dashed line).
The other parameters are chosen to be $E_z=24\, \mu\text{eV}$, $g\mu_B b_x=2\, \mu\text{eV}$, $\Omega_c/2\pi=500\, \text{MHz}$, $\gamma_{1,c}/2\pi=18\, \text{MHz}$, $\gamma_{\phi}/2\pi=600\, \text{MHz}$, and $\gamma_{M}/2\pi=2\, \text{MHz}$.
}
\end{figure}


In this section we calculate the spin Rabi frequency and the 
single-qubit average gate fidelity
for a general detuning $\varepsilon$ and study the crossover from the molecular or DQD regime ($\varepsilon=0$) to the SQD regime with the electron strongly localized in the left or right QD ($|\varepsilon|\gg 2t_c$).
An electron trapped in a detuned DQD, with energy detuning $\varepsilon$ between the left and the right QDs, 
forms charge states separated by an energy $\Omega=\sqrt{\varepsilon^2+4t_c^2}$.
The detuning
reduces the 
off-diagonal matrix elements of the transition dipole moment operator in the eigenbasis
 resulting in a Rabi frequency $\Omega'_c=\Omega_c \cos{\theta}$, where we  have introduced the  orbital angle  $\theta=\arctan{(\varepsilon/2t_c)}$, and incorporates 
diagonal matrix elements.
With a magnetic field profile as explained above,  the model Hamiltonian reads~\cite{Benito2017}
\begin{equation}
H_0=\frac{\Omega}{2}\tilde{\tau}_z+\frac{E_z}{2}\tilde{\sigma}_z-\frac{g\mu_B b_x  \tilde{\sigma}_x}{2}\left(\cos{\theta}\tilde{\tau}_x-\sin{\theta}\tilde{\tau}_z\right) \ . \label{eq:H2}
\end{equation}
The eigenenergies, labelled as $E_{0,\dots,3}$ read
$2E_{3(2)}=-2E_{0(1)}=\sqrt{\left(\Omega\pm b\right)^2+(g \mu_B b_x \cos\theta)^2}$, with
 $b=\sqrt{E_z^2+(g\mu_B b_x \sin\theta)^2}$,
and all the off-diagonal matrix elements of the electric dipole moment operator in the eigenbasis are non-zero. Therefore all the transitions can be addressed electrically,  as shown in Fig.~\ref{fig:figure1}(b) via colored arrows.
The Rabi frequencies for the transitions involving the lower energy states  are (see Appendix~\ref{sec:full-dipole}) 
$\Omega_{\sigma}=\Omega'_c \cos\Phi\sin\bar{\phi}$, and  $\Omega_{\tau}=\Omega'_c \cos\Phi\cos\bar{\phi}$,
where  the angle $\Phi=\arctan{\left(b_x \sin\theta/B_z\right)}$ describes an orbital-dependent spin rotation, $\phi_{\pm}=\arctan{\left[g\mu_B b_x \cos{\theta}/\left(\Omega\pm b\right)\right]}$ ($\phi_{\pm}\in [0,\pi]$)  generalize the spin-orbit mixing angles, and $\bar{\phi}=\left(\phi_++\phi_-\right)/2$.
%

Analogously to the previous section, we define the  spin qubit  as $s=\tau$ ($s=\sigma$) for $\Omega<E_z$ ($\Omega>E_z$), i.e., as the ground state and the second (first) excited state. As expected, the  spin qubit Rabi frequency  is reduced as $\varepsilon$ increases.
The 
expansion of $\Omega_s$ for small $b_x$ ($g\mu_B b_x\ll |\Omega-E_z|$)  yields
\begin{equation}
\Omega_s  =   2t_c g\mu_B b_x \Omega'_c  /|\Omega^2-E_z^2|+\mathcal{O}(b_x^3)  \label{eq:gs-eps} \ ,
\end{equation}
generalizing Eq.~\eqref{eq:gc-expansion-eps0} to $\varepsilon\neq0$.
In  Fig.~\ref{fig:figure2}(a), we plot the ratio $\Omega_s/\Omega_c$ as a function of $\varepsilon$  for  tunnel coupling $t_c=15\, \mu\text{eV}$ and fixed magnetic field profile, $E_z=24\, \mu\text{eV}$ and $g\mu_B b_x=2\, \mu\text{eV}$. 
As expected, for a given amplitude of the applied electric field the Rabi frequency is larger at zero detuning, which implies that at $\varepsilon \approx0$ one can drive Rabi oscillations at a given frequency with less power consumption than for finite detuning; see Appendix~\ref{sec:SQD}.
 The direct phonon-induced spin relaxation rate for small $b_x$ reads 
\begin{equation}
\gamma_{1,s}=\gamma_{1,c} (2t_c/\Omega)^2 \left[2t_c g\mu_B b_x/(\Omega^2-E_z^2)\right]^2 +\mathcal{O}(b_x^4) \ .
\end{equation}
In this detuned situation, 
the second excited state can also decay to the first excited state via phonon emission, which opens another spin relaxation channel for the case $E_z>\Omega$  (see Appendix ~\ref{sec:full-dipole}). However, the corresponding decay rate is lower than $\gamma_{1,c}$ due to the smaller energy gap between these two states and it can be neglected  for the relevant parameters. 
Moreover the  low-frequency charge fluctuations (with magnitude $\gamma_{\phi}$) induce pure dephasing with rates proportional to the first derivative of the transition frequencies with respect to $\varepsilon$,
\begin{align}
\gamma_{\phi,\tau(\sigma)}^{(1)}=\frac{\gamma_{\phi}\cos{\theta}}{2} &\left\{\tan{\theta} \left(\cos{\phi_+}\pm\cos{\phi_-}\right) \right.\nonumber\\  
&\left. +\sin{\Phi}\left(\sin{\phi_+}\mp\sin{\phi_-}\right)\right\}  \label{eq:gamma-phi-1}
\end{align}
 (see Appendix ~\ref{sec:full-dipole}), 
which yields 
\begin{equation}
\gamma_{\phi,s}^{(1)}=\frac{\gamma_{\phi}|\varepsilon|  }{E_z} \left[2t_c g\mu_B b_x/(\Omega^2-E_z^2)\right]^2 +\mathcal{O}(b_x^4) \, . \label{eq:gammasphi}
\end{equation}
The second order contribution  to spin dephasing  is proportional to the second derivatives of the transition frequencies, as 
calculated from second order perturbation theory~\cite{Cottet_thesis,Chirolli2008,Russ2015}. The full   expression for this spin contribution is given  in Appendix~\ref{sec:full-dipole}. Including terms to  lowest order in $b_x$, we find
\begin{equation}
\gamma_{\phi,s}^{(2)}=\frac{ \gamma_{\phi}^2 }{E_z} \left[\frac{2t_c g\mu_B b_x}{(\Omega^2-E_z^2)}\right]^2\left[1-\frac{4\varepsilon^2}{\Omega^2-E_z^2}\right]+\mathcal{O}(b_x^4) \ . \label{eq:gammaphi2-expansion}
\end{equation}
Finally, the  dephasing rates corresponding to quasistatic magnetic noise are 
given in
 Appendix \ref{sec:magnetic-noise} and accounting for  terms to lowest order in $b_x$, we find
 \begin{align}
\gamma_{M,s}= \gamma_{M}&\frac{\sqrt{2\epsilon^2+4t_c^2}}{\Omega} \left[1-\frac{(g\mu_B b_x\varepsilon)^2}{2E_z^2\Omega^2}\right. \nonumber\\
&\left.
-\frac{(g\mu_B b_x)^2t_c^2(\Omega^2+E_z^2)}{(2t_c^2+\epsilon^2)(\Omega^2-E_z^2)^2}\right]+\mathcal{O}(b_x^4)  \label{eq:magnetic-noise-expansion}\ .
\end{align} 
%

%
In Fig.~\ref{fig:figure2}(b), we show 
the single-qubit average gate fidelity
%
as a function of $\varepsilon$ and $t_c$, calculated by averaging the $X_{\pi}$ average gate fidelity
in the presence of Gaussian distributed noise with standard deviation
given by the total magnitude of the low-frequency noise, $\text{Var}(\delta)=2\left({\gamma_{\phi,s}^{(1)}}^2+{\gamma_{\phi,s}^{(2)}}^2+\gamma_{M,s}^2\right)$.
%
%
%
First, we can observe the optimal values of $t_c$ mentioned in Sec.~\ref{sec:flopping-mode} and a reduction in the  fidelity when $\Omega=E_z$ (indicated by the dashed line) due to large spin-charge mixing. 
Moreover, we  can see the detrimental effect of working slightly away from the sweet spot ($\varepsilon=0$). The qubit not only suffers from a lower Rabi frequency but the first order charge noise contribution dominates, abruptly decreasing the average gate fidelity.

As an estimate of the number of Rabi oscillations that can be observed with high visibility in a EDSR experiment we can use the quality factor $Q$, defined as the ratio of spin Rabi frequency and decay rates
\begin{equation}
Q=\frac{2\Omega_s}{\gamma_{1,s}/2+\sqrt{{\gamma_{\phi,s}^{(1)}}^2+{\gamma_{\phi,s}^{(2)}}^2+\gamma_{M,s}^2}} \ .
\end{equation}
This expression should be viewed as an approximate interpolation between the limiting cases where
relaxation rate $\gamma_{1,s}$ or the low-frequency noise are dominating~\cite{Croot2019}.

Increasing the detuning localizes the electron more in a single QD and the flopping-mode EDSR mechanism described above may compete with other EDSR mechanisms that take place in a  SQD, via  excited orbital  or valley states~\cite{Golovach2006,Tokura2006,Kawakami2014,Hao2014,Malkoc2016,Rancic2016,Corna2018,Bourdet2018}.
%
Also in a DQD structure, if the intervalley interdot tunnel coupling~\cite{Burkard2016,Huang2017,Mi2018b} is strong compared to the valley splittings~\cite{Mi2018b},  the   effective spin Rabi frequency  will be modified. 
In this work we focus on the  micromagnet-induced flopping-mode EDSR mechanism, which dominates if the excited orbital and valley energy splittings are large enough.
For a discussion of the interplay between micromagnet-induced SQD and flopping-mode EDSR mechanisms we refer the reader to Appendix~\ref{sec:SQD}.

In more realistic setups, where the micromagnet  stray field is not perfectly aligned with the DQD, 
there can be  magnetic field gradients  in the $z$ direction (longitudinal) and  a finite average  field in the  $x$ direction (transverse). 
Given the  importance of the protection against charge fluctuations,  we investigate the sweet spot  behavior using a  more general model in the following section.
%


\section{Flopping-mode charge noise sweet spots\label{sec:sweet-spots}}

In this section, we examine the optimal working points for flopping-mode spin qubit EDSR operation.
For the model used in Sec.~\ref{sec:crossover}, the zero detuning point constitutes a first order sweet spot with respect to fluctuations in the detuning, since the qubit energy is  insensitive to $\varepsilon$ variations to first order. In this case, it is important  to account for the second order contribution to qubit dephasing which, as mentioned above, is related to the second derivative of the qubit energy with respect to the detuning.
The  micromagnet could be designed to induce a longitudinal magnetic field  gradient between the left and the right QDs with the aim of obtaining a different spin resonance frequency depending on the electron position. Fabrication misalignments can also give rise to both longitudinal gradients and overall transverse magnetic fields~\cite{Samkharadze2018,Croot2019,Borjans2019}, i.e., the magnetic field components in the right and left QD positions may be $B^{(L,R)}_z=B_z\pm b_z$ and $B^{(L,R)}_x=B_x\pm b_x$, where $B_z\gg B_x,b_x,b_z$. Via a rotation of the spin quantization axis, given by the small angle $\zeta=\arctan{\left(B_x/B_z\right)}$, it is always possible to rewrite the latter as $B^{(L,R)}_{z'}=\sqrt{B_z^2+B_x^2}\pm b_{z'}$ and $B^{(L,R)}_{x'}=\pm b_{x'}$, with
\begin{align}
b_{z'}&=b_z \cos\zeta+b_x \sin\zeta \ ,\\
b_{x'}&=b_x \cos\zeta-b_z \sin\zeta \ ,
\end{align}
therefore  a model containing a homogeneous field and two gradients  is sufficient. In the following we 
work in a rotated coordinate system and
rename the variables 
as
$\sqrt{B_z^2+B_x^2}\rightarrow B_z$, $b_{x'}\rightarrow b_x$ and  $b_{z'}\rightarrow b_z$.
This allows us to use the 
model Hamiltonian in Eq.~\eqref{eq:H2}, with a homogeneous field $B_z$  and a  transverse inhomogeneous component $b_x$, and add a term  accounting for the  longitudinal gradient  ($\pm b_z$ in the left/right QD),
\begin{equation}
H=H_0-\frac{g\mu_B b_z \tilde{\sigma}_z}{2}\left(\cos{\theta}\tilde{\tau}_x-\sin{\theta}\tilde{\tau}_z\right)  \ . \label{eq:total-H}
\end{equation} 
Note that the relative values of $b_x$ and $b_z$ can be controlled  via the direction of the external magnetic field~\cite{Borjans2019}.

\begin{figure}
\includegraphics{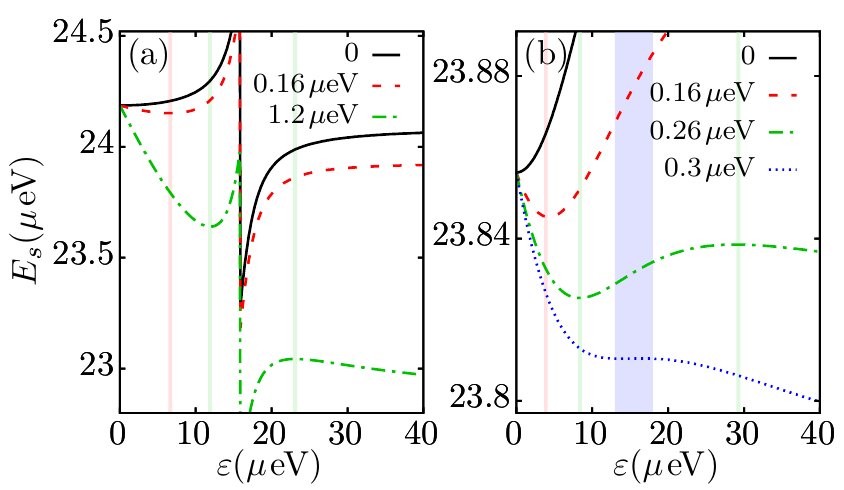}

\protect\caption{\label{fig:figure3} 
Spin qubit energy splitting $E_s$ as a function of the detuning $\varepsilon$, for various values of the longitudinal gradient field ($g\mu_B b_z$), as indicated, increasing from top to bottom. 
The interdot tunnel splitting amounts to
 (a) $2t_c=18\,\mu\text{eV}$ and   (b) $2t_c=30\,\mu\text{eV}$, while the homogeneous field Zeeman energy is $E_z=24\,\mu\text{eV}$ and the transverse inhomogeneous component is $g\mu_B b_x=2\,\mu\text{eV}$.
The thin shaded areas indicate first order sweet spots for the corresponding color  line and  the wide blue shaded area in (b) indicates the region around the second order sweet spot for $g\mu_B b_z=0.3\,\mu\text{eV}$. The discontinuity in (a) occurs at $\Omega=E_z$ due to 
a level crossing of the upper qubit state.
}
\end{figure}

For simplicity we  analyze first this model in the limit of small  inhomogeneous fields, $g\mu_B b_{x,z}\ll |\Omega-E_z|$.
While the  transverse gradient corrects  the spin qubit energy splitting $E_s$  (from the value $E_s=E_z$ for $b_{x,z}=0$) to second order, the longitudinal gradient  has an effect to first order, leading to
\begin{equation}
 E_s \simeq E_z-\frac{E_z^2-\varepsilon^2}{2E_z(\Omega^2-E_z^2)}(g\mu_B b_x)^2-\frac{\varepsilon}{\Omega} g\mu_B b_z \ .
\label{eq:energy}
\end{equation} 
From this simplified expression, we can explore the existence of first order sweet spots.
Unless $b_z=0$,  the spin qubit does not have a first order sweet spot at zero detuning. For  an arbitrary value of $t_c$,
if $b_z<b_x^2/B_z$ the spin qubit should be operated at a first order sweet spot slightly shifted from zero detuning (see below). 
For a larger longitudinal gradient, $b_z>b_x^2/B_z$, there are two first order sweet spots for a given value of tunnel splitting below the Zeeman energy,  i.e., $2t_c<E_z$.
For larger tunnel splitting, $2E_z>2t_c>E_z$, there are also two first order sweet spots  if 
\begin{equation}
\frac{b_x^2}{B_z}<b_z<b_z^{0}=\frac{3\sqrt{3}t_c^4}{E_z(4t_c^2-E_z^2)^{3/2}}\frac{b_x^2}{B_z}
\end{equation}
and none otherwise.

In Fig.~\ref{fig:figure3}, the exact spin qubit energy splitting $E_s$, calculated from the eigenenergies of the Hamiltonian~\eqref{eq:total-H}, is shown as a function of the DQD detuning $\varepsilon$ for different  values of $b_z$. 
For negative values of $b_z$ the sweet spots will occur at negative values of $\varepsilon$.
The panels (a) and (b) represent a generic case with tunnel splitting below and above the Zeeman energy, respectively.
The black (solid) lines are for $b_z=0$ and the red (dashed) lines correspond to $b_z<b_x^2/B_z$, showing therefore one first order sweet spot in both panels (a) and (b). 
In Fig.~\ref{fig:figure3}(a), since $2t_c<E_z$, we expect two first order sweet spots for large enough values of longitudinal gradient, which can be seen in the green (dash-dotted) line. 
In Fig.~\ref{fig:figure3}(b), we analyze a case with $2E_z>2t_c>E_z$. The green (dash-dotted) line corresponds to
 the intermediate region of two first order sweet spots, $b_x^2/B_z<b_z<b_z^{0}$. Finally, the blue (dotted) line is obtained  for $b_z\sim b_z^{0}$.  At this point,  $E_s$ becomes very flat, which would protect the qubit even to higher order from  fluctuations in the detuning. 

\begin{figure}
\includegraphics{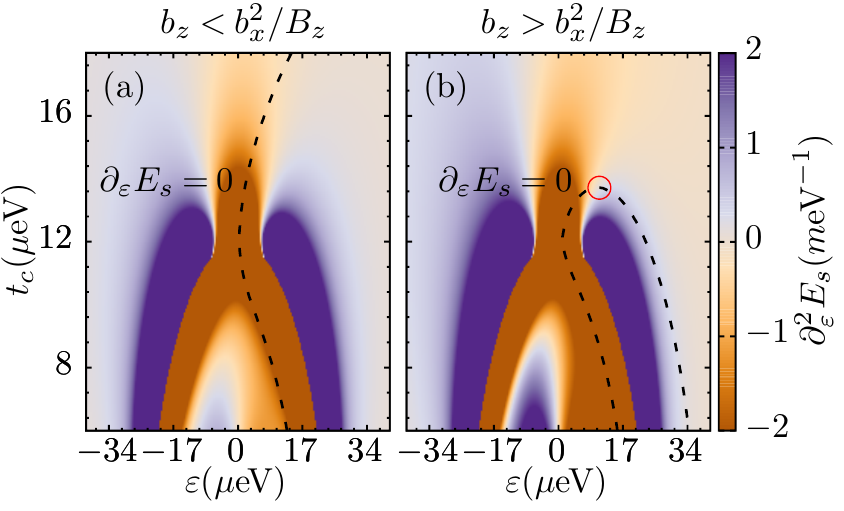}

\protect\caption{\label{fig:figure4} 
Second derivative $\partial^2_{\varepsilon}E_s$ of the spin qubit energy splitting with  respect to the detuning $\varepsilon$ as a  function of $t_c$ and $\varepsilon$ for (a) $g\mu_B b_z=0.16\, \mu\text{eV}$ and (b) $g\mu_B b_z=0.5\, \mu\text{eV}$. The black dashed lines indicate the first order sweet spot positions and the  circle in panel (b) indicates the position of the second order sweet spot. The homogeneous field Zeeman energy is $E_z=24\,\mu\text{eV}$ and the transverse inhomogeneous component is $g\mu_B b_x=2\,\mu\text{eV}$.
}
\end{figure}

 To confirm this, we show in Fig.~\ref{fig:figure4} the second derivative of the spin qubit energy splitting with respect to detuning. 
In panel (a) $b_z<b_x^2/B_z$, while in panel (b) $b_z>b_x^2/B_z$. The superimposed
black dashed line indicates  the position of the first order sweet spots. 
In Fig.~\ref{fig:figure4}(a), 
the value of the second derivative along the expected first order sweet spot (black dashed line) does not change significantly. 
Increasing the value of $b_z$ can give rise to a situation as shown in Fig.~\ref{fig:figure4}(b), where the line indicating the position of the first order sweet spot (black dashed line)
crosses the line of  zero second derivative, allowing for a second order sweet spot and a qubit  protected against charge noise up to second order. 
 %
%

The longitudinal magnetic field gradient may also influence the electric dipole moment operator and therefore the Rabi frequencies of the different transitions.
In  Appendix~\ref{sec:long-gradient} we treat the transverse component  $b_x$ perturbatively and calculate 
the correction of the spin Rabi frequency due to the longitudinal
magnetic field gradient,
\begin{equation}
\Omega_{s}\simeq \Omega'_c \frac{2t_c g\mu_B b_x}{|\Omega^2-E_z^2|}\left[1+\frac{\varepsilon b_z}{\Omega B_z}\right] \ ,
\label{eq:gsbz}
\end{equation}
i.e., $b_z\ll B_z$ incorporates  a small correction.
This means that $b_z$ does not have a noticeable effect on the spin Rabi frequency and the phonon induced spin dephasing rate, but it  strongly affects the pure spin dephasing rate due to charge fluctuations via a drastic modification of the qubit energy detuning dependence, as  shown in Figs.~\ref{fig:figure3} and~\ref{fig:figure4}.

\begin{figure}
\includegraphics{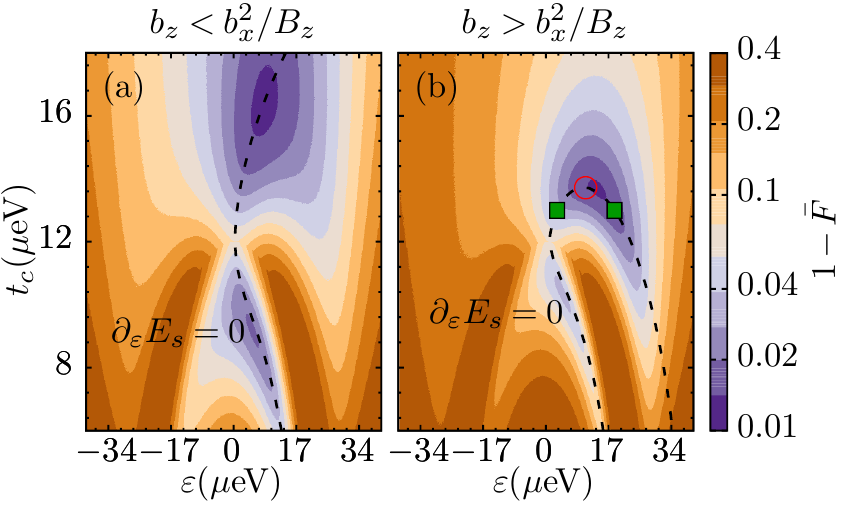}

\protect\caption{\label{fig:figure5} 
%
Single-qubit average gate infidelity $1-\bar{F}$
as a function of detuning $\varepsilon$  and   interdot tunnel coupling $t_c$ for (a) $g\mu_B b_z=0.16\, \mu\text{eV}$ and (b) $g\mu_B b_z=0.5\, \mu\text{eV}$. 
The homogeneous field Zeeman energy is $E_z=24\,\mu\text{eV}$ and the transverse inhomogeneous component is $g\mu_B b_x=2\,\mu\text{eV}$.
The other parameters are chosen to be $\Omega_c/2\pi=500\, \text{MHz}$, $\gamma_{1,c}/2\pi=18\, \text{MHz}$, $\gamma_{\phi}/2\pi=600\, \text{MHz}$, and $\gamma_{M}/2\pi=2\, \text{MHz}$.
The black dashed lines indicate the first order sweet spot positions. In panel (b) the squares mark  the position of the first order sweet spots for $t_c=13\,\mu\text{eV}$ and the  circle  indicates the position of the second order sweet spot.}
\end{figure}

To examine the overall 
performance of the qubit in different regimes, we show in Fig.~\ref{fig:figure5} the 
%
single-qubit average gate fidelity
as a function of $\varepsilon$ and $t_c$.
The charge noise induced spin dephasing rate has been calculated numerically from the derivatives of the spin qubit energy splitting $E_{s}$ with respect to detuning $\varepsilon$. The effect of the small longitudinal gradient   on the spin Rabi frequency, the phonon induced spin relaxation rate and the magnetic noise induced rate is very small, therefore we have neglected it here. Since we have assumed that the pure  dephasing rate induced by charge noise fluctuations is the dominant source of decoherence, the condition for the best quality qubit coincides with the position of the first order sweet spots, which, as opposed to the case with $b_z=0$ shown in Fig.~\ref{fig:figure2}, does not occur at $\varepsilon=0$. 
Although for a fixed tunnel coupling $t_c$ the two first order sweet spots exhibit high single-qubit average gate fidelity, their properties are very different. For example, for $t_c=13\,\mu\text{eV}$ the spin Rabi frequency at the sweet spot at $\epsilon=3.1\,\mu\text{eV}$ is four times larger than at the one at $\epsilon=18.6\,\mu\text{eV}$ (these two first-order sweet spots are indicated by squares in Fig.~\ref{fig:figure5}(b)), but the phonon-induced relaxation rate and the charge noise dephasing rates are also 16 and 9 times higher, respectively. The first order sweet spot situated at larger detuning could therefore serve as idle point, while the one at lower detuning is used as operating point.
Finally, as shown in Fig.~\ref{fig:figure5}(b),  an even larger average gate fidelity can be achieved by operating close to  the second order sweet spot.  Note that the best  fidelity does not correspond exactly to the second order sweet spot, since phonon relaxation and nuclear spin induced dephasing are also present.

\section{Conclusions}
\label{sec:conclusions}

The flopping-mode configuration is shown to be useful not only for achieving a strong coupling 
between cavity photons and single spins
~\cite{Mi2018,Samkharadze2018,Cubaynes2019}, but also for coherent electrical spin manipulation.
We have analyzed  the variation of the performance of the flopping-mode EDSR method from the symmetric ($\varepsilon=0$) DQD to the highly biased ($|\varepsilon|\gg 2t_c$) SQD regime.
Importantly, the applied power of the electric field necessary  to obtain a given Rabi frequency will be 
reduced by orders of magnitude
by working  in the DQD regime.
This efficient single spin manipulation implemented in silicon QDs would constitute a fundamental step towards a fully electrically 
controllable quantum processing architecture for spin qubits, a platform which already benefits from mature silicon processing technology.

Given the presence of  environmental charge noise  in typical QD devices, it is important to know the position of the exact first order sweet spot, which can be shifted a few  $\mu$eVs away from zero detuning in the presence of a longitudinal magnetic field gradient.
Interestingly, it is also possible to find two first order sweet spots for the same value of tunnel coupling, with different Rabi frequency and decoherence rate, which 
could be potentially exploited for different steps of qubit manipulation.
Finally, we predict the existence of
second order sweet spots, 
where the qubit is insensitive to electrical fluctuations up to second order.
%

\begin{acknowledgments}
\textit{Acknowledgments.---}
This work has been supported by the Army Research Office grant W911NF-15-1-0149 and the DFG through SFB767. 
We would also like to acknowledge B. D'Anjou and M.
Russ for helpful discussions.
\end{acknowledgments}

\appendix

\section{Electric dipole moment and dephasing 
\label{sec:full-dipole}}

In this Appendix we calculate the  Rabi frequencies for the different transitions in the flopping-mode spin qubit, the phonon-induced spin relaxation rates and the pure dephasing rates due to low-frequency electrical fluctuations in the DQD detuning.
In Eq.~\eqref{eq:H0-diagonal} we have expressed the electric dipole moment operator in the eigenbasis of Eq.~\eqref{eq:H0-eps0}, which is the model Hamiltonian for $\varepsilon=0$ and $b_z=0$.
For detuned QDs ($\varepsilon\neq0$), we can write the electric dipole moment in the eigenbasis of the Hamiltonian in Eq.~\eqref{eq:H2} and find that the electric field couples to all possible electronic transitions, as shown in Fig.~\ref{fig:figure1}(b), since the electric dipole moment operator has the form
$p=e d \cos{\theta} (\mathcal{T}+\mathcal{Z}/2)$, 
with the off-diagonal  component
\begin{align}
\mathcal{T}=&
-\cos{\Phi}\cos{\bar{\phi}} \tau_x  +\cos{\Phi}\sin{\bar{\phi}}\sigma_x\tau_z \label{eq:transversal}\\
&+\left(\sin{\Phi}\cos{\phi_-}+\tan{\theta}\sin{\phi_-}\right) (\sigma_+\tau_-+h.c.) \nonumber\\
&-\left(\sin{\Phi}\cos{\phi_+}-\tan{\theta}\sin{\phi_+}\right) (\sigma_+\tau_++h.c.) \ ,\nonumber
\end{align}
and the diagonal component 
\begin{align}
\mathcal{Z}=
&\left\{\tan{\theta} \left(\cos{\phi_+}+\cos{\phi_-}\right)\right. \nonumber \\ &\left.+\sin{\Phi}\left(\sin{\phi_+}-\sin{\phi_-}\right)\right\}\tau_z
\nonumber\\
+&\left\{\tan{\theta}\left(\cos{\phi_+}-\cos\phi_-\right)\right.\nonumber\\ &\left.+\sin{\Phi}\left(\sin{\phi_+}+\sin{\phi_-}\right)\right\}\sigma_z\  . \label{eq:longitudinal}
\end{align}
The first terms in the off-diagonal component determine the Rabi frequencies $\Omega_{\tau(\sigma)}$  and the direct phonon relaxation rates $\gamma_{1,\tau(\sigma)}$ given in Sec.~\ref{sec:crossover}. The term in the second line of Eq.~\eqref{eq:transversal} corresponds to transitions between the first and second excited states, and it opens a new channel for spin relaxation in the case $E_z>\Omega$. We have neglected this channel here because the corresponding phonon emission rate is suppressed by the small energy gap between these two states for the relevant parameter regimes.

The electrical fluctuations also couple to the system via the electric dipole moment. If the amplitude $\delta_{\varepsilon}$ and frequency of these fluctuations is small, we can calculate the spin qubit dephasing rate by treating them within time-independent perturbation theory~\cite{Cottet_thesis,Chirolli2008,Russ2015}, obtaining the dephasing Hamiltonian
\begin{equation}
H_{\delta_{\varepsilon}}=\sum_{\eta=\tau,\sigma}\left(\frac{\partial E_{\eta}}{\partial \varepsilon}\delta_{\varepsilon}+\frac{1}{2}\frac{\partial^2 E_{\eta}}{\partial \varepsilon^2}\delta_{\varepsilon}^2\right)\frac{\eta_z}{2} \ ,
\end{equation}
where the first order contribution relates directly to the  diagonal components in Eq.~\eqref{eq:longitudinal}, since
\begin{align}
\frac{\partial E_{\tau(\sigma)}}{\partial \varepsilon}=\frac{\cos{\theta}}{2} &\left\{\tan{\theta} \left(\cos{\phi_+}\pm\cos{\phi_-}\right) \right.\nonumber\\  
&\left. +\sin{\Phi}\left(\sin{\phi_+}\mp\sin{\phi_-}\right)\right\}  \ . \label{eq:gamma-phi-1-app}
\end{align} 
and all the terms of the off-diagonal component Eq.~\eqref{eq:transversal} contribute to second order~\cite{Russ2015}. 
More precisely, the second derivatives read
 \begin{align}
\frac{\partial^2 E_{\tau(\sigma)}}{\partial \varepsilon^2}&=\cos^2{\theta}\left\{\frac{\cos^2{\Phi}\cos^2{\bar{\phi}}}{E_{\tau(\sigma)}}\right.\nonumber\\&\left.+\frac{\left(\sin{\Phi}\cos{\phi_+}-\tan{\theta}\sin{\phi_+}\right) ^2}{2(E_{\tau}+E_{\sigma})}\right.\nonumber\\&\left.\pm \frac{\left(\sin{\Phi}\cos{\phi_-}+\tan{\theta}\sin{\phi_-}\right)^2}{2(E_{\tau}-E_{\sigma})}\right\} \ .
\end{align}

Assuming Gaussian distributed  low frequency noise leads to a Gaussian decay of coherence $\propto e^{-\Gamma_{\phi}^2 t^2}$ with the total pure spin dephasing rate related to the variance of the noise function
\begin{align}
\Gamma_{\phi}&=\left[\text{Var}\left( \frac{\partial E_{s}}{\partial \varepsilon}\delta_{\varepsilon}+\frac{1}{2}\frac{\partial^2 E_{s}}{\partial \varepsilon^2} \delta_{\varepsilon}^2\right)/2 \right]^{1/2} \nonumber \\
&=\left[{\gamma_{\phi,s}^{(1)}}^2+{\gamma_{\phi,s}^{(2)}}^2\right]^{1/2} \ ,
\end{align}
where 
$\gamma_{\phi,s}^{(1)}=\gamma_{\phi} \frac{\partial_{\varepsilon} E_{s}}{\partial \varepsilon}$,
$\gamma_{\phi,s}^{(2)}=\gamma_{\phi}^2 \frac{\partial_{\varepsilon}^2 E_{s}}{\partial \varepsilon^2}$,
and $\gamma_{\phi}=\sigma_{\varepsilon}/\sqrt{2}$, where $\sigma_{\varepsilon}$ is the standard deviation of the fluctuations $\delta_{\varepsilon}$.
%

\section{Quasistatic magnetic noise \label{sec:magnetic-noise}}

In this Appendix we calculate the dephasing rate of the flopping-mode spin qubit due to hyperfine interaction  with the nuclear spins. For this we use the quasistatic approximation~\cite{Taylor2007}, which assumes that the fluctuations in the Overhauser field  occur in a time scale much longer than the system dynamics. Then we treat the noise Hamiltonian term 
\begin{equation}
\tilde{V}=\xi_L(t) \tilde{\sigma}_z(1+\tilde{\tau}_z)/2+\xi_R(t) \tilde{\sigma}_z(1-\tilde{\tau}_z)/2 \ , \label{eq:noise-ori}
\end{equation}
with two random variables for the noise in the left and right QDs,
to first order in  time-independent perturbation theory. First we transform Eq.~\eqref{eq:noise-ori} into the eigenbasis of Eq.~\eqref{eq:H2}, obtaining the diagonal component
\begin{align}
\mathcal{Z}=\frac{\xi_+\cos{\Phi}}{4}\left\{\left(\cos\phi_+-\cos\phi_-\right) \tau_z \right. & \nonumber\\
\left. +\left(\cos\phi_++\cos\phi_-\right)\sigma_z\right\}&\nonumber \\
+\frac{\xi_-\cos{\Phi}\sin\theta}{2}\sigma_z\tau_z&\ , \label{eq:noise}
\end{align}
where $\xi_{\pm}=\xi_L(t)\pm\xi_R(t)$. 

If we assume now Gaussian distributions with zero mean value and $\sigma_M^2=\text{Var}\left( \xi_R(t)\right)=\text{Var}\left( \xi_L(t)\right)$, the coherences decay as $\propto e^{-(\gamma_{M,\sigma(\tau)} t)^2}$, with the  dephasing rates due to nuclear spins
\begin{equation}
\gamma_{M,\sigma(\tau)}=\frac{\gamma_{M}  \cos{\Phi}}{2}
\sqrt{\left(\cos\phi_+\pm\cos\phi_-\right)^2+4\sin^2\theta}
 \ ,
\end{equation}
where $\gamma_M=\sigma_M$,
whose expansion to lowest order in $b_x$ yields Eq.~\eqref{eq:magnetic-noise-expansion}. 


\section{Single-qubit average gate fidelity
\label{sec:average-fidelity}}

We determine the quality of the quantum gate, represented by the operator $\cal{E}$, via the average fidelity
$\bar{F} = \overline{\langle \psi|
  \cal{E} [|\psi _{\rm i} \rangle]|\psi\rangle}$, which compares the targeted  pure state  $\ket{\psi}$ and the obtained mixed state density matrix $\cal{E} [|\psi _{\rm i}
 \rangle]$,
 averaged over all possible pure input
 states $|\psi _{\rm i}\rangle$.
 In this case the real quantum gate is determined by the simple two-level system master equation
 \begin{equation}
 \dot{\rho}=-i\left[\frac{\delta}{2}\sigma_z,\rho\right]+\frac{\gamma_{1,s}}{2}\left[2\sigma_-\rho \sigma_+-\left\{\sigma_+\sigma_- ,\rho\right\}\right]
 \end{equation}
 for the qubit density matrix $\rho$, where $\delta$ is the noise magnitude.
 
 We now calculate the entanglement fidelity $F_e$ for the gate applied to only one qubit  of a two-qubit state prepared in a maximally entangled state, since this relates to the average fidelity as $\bar{F}=(2 F_e+1)/3$~\cite{Horodecki1999}. This yields
\begin{align}
\bar{F}(\delta)&=\frac{1}{3}\left\{2+e^{-2t_g \gamma_{1,s}}\right. \label{eq:Fidelity}\\
&\left.+e^{-t_g \gamma_{1,s}}\left[\cosh{\left(t_g\gamma_{1,s}\right)}-\cosh{\left(t_g\sqrt{\gamma_{1,s}^2-\delta^2}\right) }\right]\right\} \ . \nonumber
\end{align}
Finally, since we consider only low-frequency noise, the measurable and interesting quantity is the average of this fidelity over the randomly distributed noise variable $\delta$.

\section{Low power EDSR}\label{sec:SQD}
In this Appendix, we analyze the power necessary to drive Rabi oscillations at a given frequency by taking into account both SQD and flopping-mode EDSR induced by the micromagnet. Following Refs.~\cite{Pioro-Ladriere2008,Kawakami2014}, we can complete Eq.~\eqref{eq:gs-eps} by including the SQD contribution to the Rabi frequency,
\begin{equation}
\Omega_s  \approx   \frac{e d E_{ac}}{\hbar}g\mu_B b_x \left(\frac{4t_c^2  }{\Omega|\Omega^2-E_z^2|}+\frac{\hbar^2}{m_e^* d^2E^2_{\text{orb}}}\right)  \label{eq:gs-eps-SQD} \ ,
\end{equation}
where 
$E_{\text{orb}}$ is the orbital energy, $E_{\text{orb}}\approx1-3\,\text{meV}$, and $m_e^*$ is the  electron effective mass. Since the drive power is proportional to the square of the electric field, $P\propto E_{ac}^2$, the power necessary to drive the spin qubit at a given Rabi frequency follows~\cite{Croot2019}
\begin{equation}
P  \propto \Omega_s^2 \left[ \frac{e d}{\hbar}g\mu_B b_x \left(\frac{4t_c^2  }{\Omega|\Omega^2-E_z^2|}+\frac{\hbar^2}{m_e^* d^2E^2_{\text{orb}}}\right) \right]^{-2} \label{eq:power} \ .
\end{equation}

\section{Effect of $b_z$ on the spin Rabi frequency
\label{sec:long-gradient}}

In this Appendix
 we investigate the effect of
a longitudinal magnetic field  gradient  on the flopping-mode Rabi frequencies.
Since $b_z$ is the difference in longitudinal magnetic field  between the left and the right QDs, it can be seen as a detuning parameter (similar to $\varepsilon$) that depends on the spin, therefore its effect can be included in the form of a  spin-dependent orbital basis transformation,
\begin{align}
|+',\sigma\rangle=&\cos{(\theta_{\sigma}/2)}|+,\sigma \rangle-\sin{(\theta_{\sigma}/2)}|-,\sigma\rangle\nonumber  \ ,\\
|-',\sigma\rangle=&\sin{(\theta_{\sigma}/2)}|+,\sigma \rangle+\cos{(\theta_{\sigma}/2)}|-,\sigma\rangle
\ ,
\end{align}
 with orbital angles
$\theta_{\uparrow(\downarrow)}=\arctan{\left[(\varepsilon\pm g\mu_B b_z)/2t_c\right]}$ and orbital energies 
$\Omega_{\uparrow(\downarrow)}=\sqrt{(\varepsilon\pm  g\mu_B b_z)^2+4t_c^2}$, instead of the $\theta$ and $\Omega$ used in Sec.~\ref{sec:crossover}.
With this, we can treat $b_x$ perturbatively and find the spin Rabi frequency 
\begin{equation}
\Omega_s \simeq    2t_c g\mu_B b_x \Omega_c \cos{\bar{\theta}}  \frac{E_z/\left[E_z-(\Omega_{\uparrow}-\Omega_{\downarrow})/2\right]}{(\Omega_{\uparrow}+\Omega_{\downarrow})^2/4-E_z^2} \ ,
\end{equation}
that generalizes the result in Eq.~\eqref{eq:gs-eps}.
Here,     $\bar{\theta}=(\theta_{\uparrow}+\theta_{\downarrow})/2$.
Finally, expanding to lowest order in  $b_z$, this simplifies to Eq.~\eqref{eq:gsbz}.

\bibliography{../../../postdocreferences.bib}

\end{document}